\documentclass[a4paper, 10pt]{IEEEtran}
\usepackage{graphicx}
\usepackage{subfigure}
\usepackage[cmex10]{amsmath}
\usepackage{amssymb}
\usepackage[small,bf,up]{caption}

\newcommand{\dg}{\ensuremath{^\circ}}

\newcommand{\bV}{\mbox{\boldmath$V$}}
\renewcommand{\thetable}{\arabic{table}}
\begin{document}
\title{Graceful Degradation of Air Traffic Operations}
\author{
  Maxime Gariel\thanks{Graduate Research Assistant, School of Aerospace Engineering,
    maxime.gariel@gatech.edu} and
  Eric Feron  \thanks{ Professor, School of Aerospace
  Engineering, feron@gatech.edu.}\\
  {\normalsize\itshape
   Georgia Institute of Technology Atlanta, GA, 30332-0150, USA}\\
 }

\date{\today}
\maketitle

\begin{abstract} The introduction of new technologies and concepts of operation in
the air transportation system
is not possible, unless they can be proven not to adversely affect the
system operation under not only nominal, but also degraded conditions. In
extreme scenarios, degraded operations due to partial or complete
technological failures should never endanger system safety. Many past
system evolutions, whether ground-based or airborne, have been based on
trial-and-error, and system safety was addressed only after a specific
event yielded dramatic or near-dramatic consequences. Future system
evolutions, however, must leverage available computation, prior knowledge
and abstract reasoning to anticipate all possible system degradations and
prove that such degradations are graceful and safe. This paper is concerned
with the graceful degradation of high-density, structured arrival traffic
against partial or complete surveillance failures. It is shown that for
equal performance requirements, some traffic configurations might be easier
to handle than others, thereby offering a quantitative perspective on these
traffic configurations' ability to ``gracefully degrade''. To support our
work, we also introduce a new conflict resolution algorithm, aimed at
solving conflicts involving many aircraft when aircraft position information is in
the process of degrading.
\end{abstract}

 \section{Introduction}

Air Traffic Management (ATM) relies on several layers of technology
supporting three essential functions: Communication, Navigation and
Surveillance (CNS). Advances of the CNS technology base can directly lead
to improved air traffic management operations. And indeed,
the air traffic management system is left with no choice but to leverage
the concurrent advent of digital communication technology, satellite-based
navigation and overall improvements of available instrumentation to adjust
its ability to handle a fast-growing traffic demand.
The resulting Next Generation air transportation systems is described in
detail by the Joint Planning and Development Office \cite{nextGen} for US
operations, and by the {SESAR Consortium} \cite{SESARdeliverable3, SESARconceptOfOperation} for European
operations. 

For example, one of the cornerstones of expanded operations is ADS-B, a
navigation and surveillance concept based on the GPS satellite positioning
system which offers the potential for giving pilots more flight autonomy
during the en-route flight phase and enabling higher and more flexible
traffic densities in terminal
areas. One of the main benefits of ADS-B is to improve navigation precision by providing an accurate position fix. 
Such technology also enables so-called trajectory-based operations
(TBOs), and strategic traffic separation management using the concept of
4-dimensional trajectories (4DTs), whereby an aircraft is able to forecast and
broadcast its intended trajectory well before its execution.

However, the obligation for the ATM system to maintain very high
reliability and safety levels implies that such new system technologies can
be implemented
only if they lead to a system with equal or better
safety characteristics than currently demonstrated. While system safety
includes the
ability for the system to operate well under nominal conditions, it is also
concerned with off-nominal system behaviors, whereby operations are
expected to still remain accident-free for all known failure modes
of the system. Failures may affect several parts of the ATM infrastructure:

\begin{itemize}
 \item \textbf{ATM computational infrastructure: } Computers and
communication systems (both ground-based and airborne) form the backbone of
the air transportation system information infrastructure. Computers are not
exempt from such failures, whether the failures involve hardware
(motherboard and wiring) or software (incomplete functional requirements or
erroneous software implementation).
\item\textbf{Communications:} A communication mishap can lead to severe
consequences. Most lately, on September $25^{th}$, 2007, a communication
failure at Memphis' Air Route Traffic Control Center (ARTCC) shut down many phone lines, radio communications
and radar coverage, completely incapacitating operations  for a period
of two hours, and despite the presence of about 200 aircraft in the center.
\item\textbf{Surveillance:} Computer or radar failures can cause
surveillance failures. For instance, in December 2000, Miami International
airport was subject to repeated radar failures when erroneous flight
information data appeared on the controllers' screen. During one such
mishap, about 125 flights were rerouted, affecting traffic all over the US.
\item \textbf{Operations:} For current operation degraded modes, there
exist backup procedures described by ICAO in  \cite{ICAO4444}. Also, issues in closely-spaced parallel runway operations is the object of several papers \cite{pritchett3, pritchett1}. An analysis of current en route air traffic control usage during special situations can be found in \cite{ATCspecialSituations}. Operation failures can lead to accidents such as  the 2002 crash over Germany between a Russian passenger jet and a cargo plane. Contradictory orders provided by the Traffic Collision Avoidance System (TCAS) and the controller lead to a collision. 
\item \textbf{Vehicles:} An intruder can enter the airspace and
consequently, jeopardize the safety of surrounding traffic. A private plane
can unintentionally  get too close to a restricted airspace such as the
vicinity of major airports. If a VFR (Visual Flight Rules) aircraft gets in
the landing path of IFR (Instrument Flight Rules) flights, this generates
an abnormal situation to be solved by the controller.
\item \textbf{Airport closure:}  An airport or part of it can be closed,
e.g for weather conditions. The traffic needs to be reorganized and
rerouted towards other airports. For instance, Bangor International Airport
is often a diversion destination when freezing rain, snow and fog close
Boston, New York or other major Northeast metropolitan centers.
\end{itemize}

In the past, such failure modes and how to recover from them have been
identified on an {\em ad hoc} basis, whereby accidents have triggered
extensive studies and redesigns of the air traffic management operations.
Extensive experience about incidents by air traffic controllers has
progressively led them to always address ``what if'' questions during
routine operations, leading them to safe operations. This system comes
complete with degraded operation and recovery procedures, such as those
described in \cite{ICAO4444}. One example of such fault-tolerant, or fail-safe
procedure concerns departure operations: A safe, collision-free path is
completely specified to the aircraft prior to take-off, in such a way that
the aircraft may follow this path safely even in the case of complete
communication failure during take-off.

ADS-B, digital communication technology and advanced automation will enhance  higher density airborne operations
in the proximity of busy airports, en-route or terminal airspace operations
with flexible routes, or cockpit-centric, decentralized conflict
management. Those future evolutions of the system  will need to be proven fully safe prior to their
introduction within the operational landscape. All failure modes will need to be identified and shown to be handled safely prior to the new procedures implementation.

Within this paper, the process by which the current or future system keeps operating safely
despite degradation of the sustaining CNS infrastructure will be called
``Graceful Degradation''. The term ``Graceful Degradation'' finds its
origins in complex computer
systems~\cite{HerlihyGracefulDegradation}. However, it can be
immediately extended to overall systems such as the National Airspace System (NAS), which include both
physical assets (airplanes) and complex information infrastructures. The
examination of concepts of operation such as NextGen in the
US~\cite{nextGen} and SESAR~\cite{SESARdeliverable3, SESARconceptOfOperation} in Europe
reveals that system safety and graceful degradation are considered open
research issues for most future operations. Tables   \ref{tab:SESARResearch} and \ref{tab:NextgenResearch} present some of
the Research and Development topics for SESAR and NextGen, respectively. In many regards, the works that most closely relate to the present paper are those devoted to traffic and airspace complexity \cite{mogford1995cca,delahaye,keumjinLee,airspaceComplexitySridhar,complexityResolutionVersteegt,ishutkina2005dat}. Indeed, these also aim at evaluating, in a broad sense, the resilience of ongoing controlled traffic against off-nominal events.

\begin{table}[ht]
\begin{center}
    \begin{tabular}{|p{0.1\linewidth}|p{0.8\linewidth}|}
    \hline
    \multicolumn{1}{|c|}{\textbf{No}} & \multicolumn{1}{|c|}{\textbf{Research topic}}\\
  \hline
\multicolumn{1}{|c|}{15} & Study of the following automation topics:
   Automated separation tools and safety, impact of automation on capacity and impact of loss of situation awareness and tools to manage exceptions associated with loss of situation  awareness.\\
    \hline
\multicolumn{1}{|c|}{16} & Evaluation of ground based de-confliction automation support tools with particular focus on how to ensure
feasible solutions with a minimum of constraints on the users’ trajectory.\\
    \hline
\multicolumn{1}{|c|}{41} &Evaluation of terminal route structure design involving alternative arrival techniques with multiple or single
merging points.\\ 
    \hline
\multicolumn{1}{|c|}{42} &Evaluation of Time Based Separation (TBS) on merging points focussing on accuracy requirements and benefits.\\
    \hline
\multicolumn{1}{|c|}{49} & Study on dynamic risk modelling and management techniques for on-line measurement of safety risk. Study on the assessment of the overall safety of the CONOPS. For now, it is not obvious that the concept’s ideas all together are “safe in principal” (as stated e.g. in Episode 3 objectives).\\
    \hline
\multicolumn{1}{|c|}{51} & Model complex scenarios of new trajectory based arrival/departure techniques plus existing SID/STAR and
also with the SID/STAR from nearby airports plus transit traffic\\
    \hline
\multicolumn{1}{|c|}{54} & Study controller acceptability of ASAS Spacing versus ASAS Separation during the organization of streams of traffic\\
    \hline
\multicolumn{1}{|c|}{62} & Development, evaluation and agreement on separation minima for each separation method included in the concept\\
\hline
    \multicolumn{1}{|c|}{64} & The new separation modes described – at least Dynamic Route Allocation, 4D Contracts and ASAS-Self Separation in mixed mode environment – shall be assessed with regard to maturity and potential performance: New separation modes shall be assessed with regard to maturity and potential performance: The robustness and stability of the various methods in the face of unexpected events (even of small magnitude) is to be investigated. \\
    \hline
       \multicolumn{1}{|c|}{80} & Elaboration of high density separation concepts and associated airspace issues in terms of detail procedures which should be then validated with a focus on feasibility.\\
    \hline
 \end{tabular}\caption{{SESAR} R\&D topics \cite{SESARconceptOfOperation}}\label{tab:SESARResearch}
\end{center}
\end{table}

\begin{table}[ht]
\begin{center}
    \begin{tabular}{|p{0.08\linewidth}|p{0.33\linewidth}|p{0.45\linewidth}|}
    \hline
    \multicolumn{1}{|c|}{\textbf{Ref}} & \multicolumn{1}{|c|}{\textbf{Line
Reference}}& \multicolumn{1}{|c|}{\textbf{Issue}}\\
    \hline
    \multicolumn{1}{|c|}{R-6} & In addition, backup functions are distributed throughout the system, and there are layers of protection to allow for graceful degradation of services in the event of automation failures. & \textbf{Develop guidance for what flexibility is allowed} in the implementation of \textbf{airborne separation management} algorithms to ensure operationally consistent results, \textbf{understanding whether variations in algorithms can result in major impacts on overall operations}.\\
    \hline
       \multicolumn{1}{|c|}{R-9} & C-ATM is the means by which flight operator objectives are balanced with overall NAS performance objectives and accomplishes many of the objectives for CM, FCM, and TM. & Super Density operations will result in many aircraft in close proximity. Consequently \textbf{an aircraft deviating from its assigned trajectory is much more likely to cause an  immediate conflict with another aircraft, and safe avoidance maneuvers may be limited or unavailable. How can super density operations be conducted safely, especially in the presence of severe weather?}\\
    \hline
\multicolumn{1}{|c|}{R-10} & C-ATM is the means by which flight operator objectives are balanced with overall NAS performance objectives and accomplishes many of the objectives for CM, FCM, and TM.& \textbf{Develop requirements for a collision avoidance system that is compatible with NextGen tactical separation?} Unless mandated otherwise, some aircraft will likely be equipped with legacy TCAS/ACAS systems which may generate unwanted alerts during normal operations. How should this be accounted for?\\
    \hline
    \multicolumn{1}{|c|}{R-28} & In all, these new kinds of flight operations \textbf{dramatically improve en route productivity and capacity} and are essential to achieving NextGen. & \textbf{If the automation fails, what is the backup plan in terms of people/procedures/automation?}\\
\hline
\multicolumn{1}{|c|}{R-44} & As illustrated in Figure \ref{fig:superDensity}, superdensity corridors handle arriving and departing traffic, while much nearby airspace remains available to other traffic.& \textbf{We do not prove today's ATC system is safe, but rely on historical data. NextGen will be required to both be safe and to demonstrate it is safe. How will safety be designed into all aspects of
NextGen and then proven?}\\ 
    \hline
\multicolumn{1}{|c|}{P-1} & In addition, backup functions are distributed throughout the system, and there are layers of protection to allow for graceful degradation of services in the event of automation failures.& \textbf{Develop policies concerning liability for delegated separation and self-separation operations.}\\
\hline
    \end{tabular}\caption{NextGen research and policies issues (Part 2)}\label{tab:NextgenResearch}
\end{center}
\end{table}

In this paper, we are interested in analyzing the conditions for the
graceful degradation of high-density traffic when the positioning system
substantially degrades. The motivation behind
our work is the widespread and growing implementation of GPS-based ADS-B as
a replacement for other, conventional navigation and surveillance
mechanisms such as secondary radar and beacon-based positioning systems.
Such development offers the potential for enabling higher traffic densities
in the vicinity of large and busy airports, enabling ``super-density operations''\cite{nextGen}. For that purpose, the remainder
of this paper is organized as follows: First, we state the
graceful  degradation problem in the context of high traffic densities and
failing navigation systems. Then, an algorithm of avoidance
under uncertainties is used to compare the ability for Miles In Trail and Free flight
configurations to manage a loss of accuracy in the positioning
system. The avoidance process under uncertainties is analyzed. The algorithm is presented in appendix.

\section{Graceful degradation of surveillance and navigation systems}

This paper focuses on the impact of surveillance and navigation system
degradation on aircraft separation requirements. Indeed, the primary
mission of the air traffic control system is to ensure safe aircraft
separation under all regular and degraded circumstances. Conventional
surveillance include radar-based technology, such as primary and secondary
radars, and ground-based beacons. New, higher-resolution surveillance
systems are enabled by satellite-based positioning systems. Such technologies may enable reduced
horizontal separation minima and therefore higher airborne aircraft densities.

We may then formulate the following questions:

\begin{enumerate}
\item{\bf Analysis: Graceful degradation sensitivity.} \emph{Consider a
given air traffic situation (consisting of a number of aircraft with given
positions, velocities and headings), what is the ``sensitivity'' of this
situation relative to a sudden degradation of the surveillance system? }
\item{\bf Design: Graceful degradation-compatible guidance.} \emph{Consider a set
of aircraft with given origins and destinations. Find a sequence of
aircraft headings such that the graceful degradation sensitivity of the
overall traffic of interest remains below a given threshold.}
\end{enumerate}

\begin{figure}[ht]
\centering
    \includegraphics[scale=0.35]{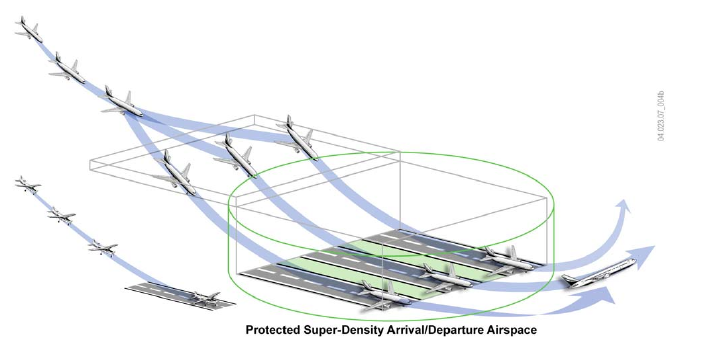}
    \caption{Super-Density operations~\cite{nextGen}}
\label{fig:superDensity}
\end{figure}

In this paper, we will be concerned with the analysis question, leaving the
design question for further research. The principle that will drive our
analysis can be sketched as follows: Considering a set of
aircraft operating under a ``high performance'' surveillance system, we
analyze whether safety can be maintained despite a failure of the
surveillance system. Assuming that failures of the surveillance system
consist of a partial loss of vehicle coverage, we will be
interested in what maneuvers will allow aircraft to remain provably
separated. Indeed,  partial or complete loss of aircraft position coverage
results in growing uncertainty about aircraft positions, in such a way that
aircraft initially close to each other may not be distinguishable from each
other shortly after the failure, unless they maneuver to augment their
physical separation. Thus, underpinning our analysis of the the ability for
a particular airborne configuration to gracefully degrade, we find the
necessity to design procedures enabling traffic to maintain provable
separation under degraded conditions. In this paper, such procedure will
consist of a novel aircraft conflict management algorithm.

Considering a planar traffic environment for simplicity, we introduce the following definitions:

\begin{enumerate}
\item {\bf Nominal operations:} \emph{ Nominal operations consist of all allowable aircraft operations when
aircraft position accuracy is high. The nominal minimum aircraft separation distance (expressed in nautical miles) will be denoted $r_0$ .}
\item {\bf Degraded operations:} \emph{Degraded operation consist of all allowable aircraft operations when aircraft position accuracy is low. The degraded minimum aircraft separation distance (expressed in nautical miles) will be denoted $r_f$, with $r_f>r_0$. }
\end{enumerate}

Radar precision is one of the main reason for deciding on specific aircraft separation standards. The uncertainty in the position seen by the controllers leads to a separation requirement that can be interpreted as a circle of avoidance around each aircraft. The circle of avoidance corresponds to the area around the aircraft where no other aircraft is allowed. Its radius is generally 2.5 NM for en-route and 1.5 NM for approach. This separation distance ensures safety if the position of the aircraft is relatively well known and regularly updated. Accurate positioning systems such as ADS-B will probably enable a reduction in allowable spacing distance \cite{powell2005}. If a failure happens, the system works in degraded mode, resulting in an increase in uncertainties on the aircraft position observed by the controller. As aircraft's position are known less accurately, the resulting radius of avoidance must be increased. The growth of the avoidance circle is limited by backup positioning system (Primary Surveillance Radars, radio...) that enable controllers to get reports on aircraft's position.  For instance, in the case of the breakdown of a radar, separation distances must be increased to procedural separation standards \cite{ICAO4444}. The position of the aircraft will be reported by the pilot to the controller by radio with a low update rate. Between updates, the position of the aircraft is not known and the uncertainty on it increases with time. 

In the remainder of this paper, we will assume that a position system failure occurs at time $t=0$. Prior to that time, nominal aircraft position accuracy will translate into a circle of avoidance with radius
$r_0$.  

\begin{figure}[ht]
\centering
 \includegraphics[scale = 0.25]{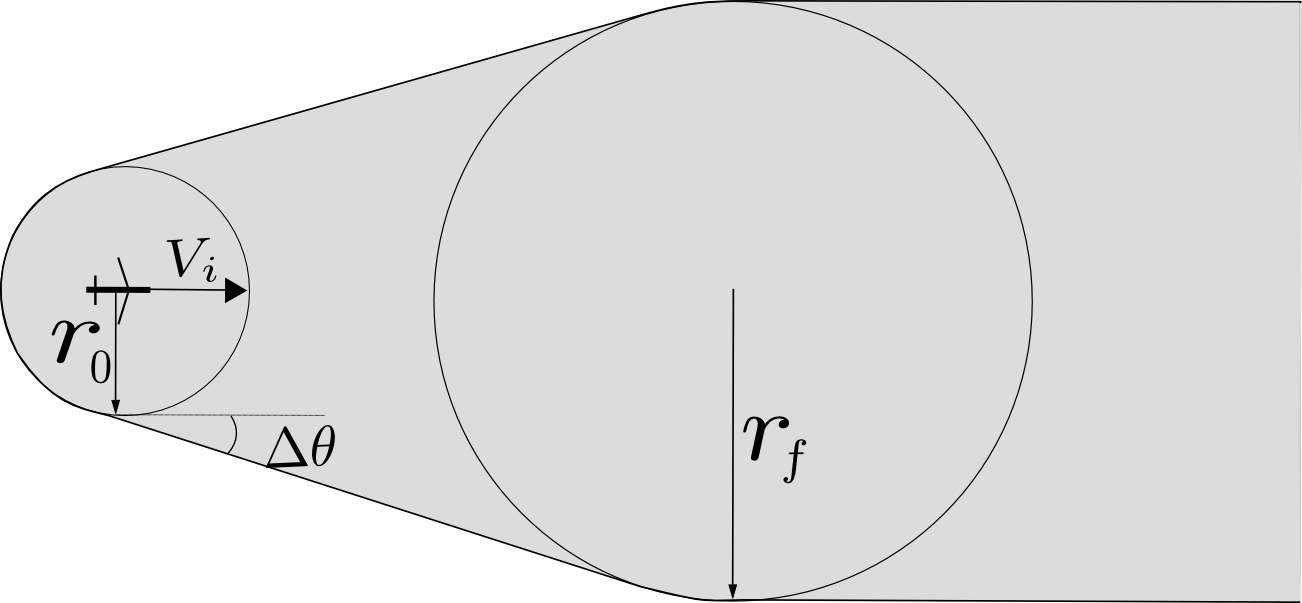}
\caption{Track of a growing circle of avoidance}\label{fig:configavoidance}
\end{figure}

For $t\geq 0$, we propose a time-varying model of position uncertainties, whereby the radius of avoidance grows from $ r_0$ to $r_f >r _0$ over a given period of time.  Figure \ref{fig:configavoidance} presents the track of a growing circle of avoidance. For instance, $r_0$ can be the radius of avoidance provided by an ADS-B positioning system, while $r_f$ can be the one provided by a Primary Surveillance Radar (PSR). In the event of an ADS-B failure, the transition from $r_0$ and $r_f$ must be eventless, in the sense that the transition should not jeopardize the safety of the overall traffic.

\begin{figure}[ht]
\centering
    \includegraphics*[scale=0.3, viewport=100 80  665 370]{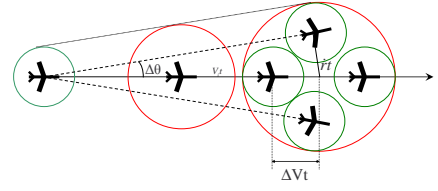}
    \caption{Uncertainty on the trajectory}
\label{fig:deltaTheta}
\end{figure}

 For each aircraft $i$, we will assume a constant growth rate $\dot r_i$, such that $r(t) = r_0 + \dot r_it$.  Such a model approximately captures the growing, but bounded uncertainty on aircraft position once the navigation system has failed. Such uncertainty might reflect the effect of uncertainties on the aircraft heading (denoted $\Delta\theta$) and on the aircraft velocity (denoted $\Delta V_i$). Figure \ref{fig:deltaTheta} shows the uncertainty on the trajectory. A simple way to connect these uncertainties to the growing avoidance radius is to write, for example,
\begin{equation}
\dot r_i = \max\{V_i\sin{\Delta\theta},  \Delta V_i\}.
\end{equation}

Note this is a conservative approximation.

\begin{figure}[ht]
\begin{center}
\subfigure[Classical conflict avoidance ]{\label{fig:classicConflict}
   \includegraphics[scale = 0.3]{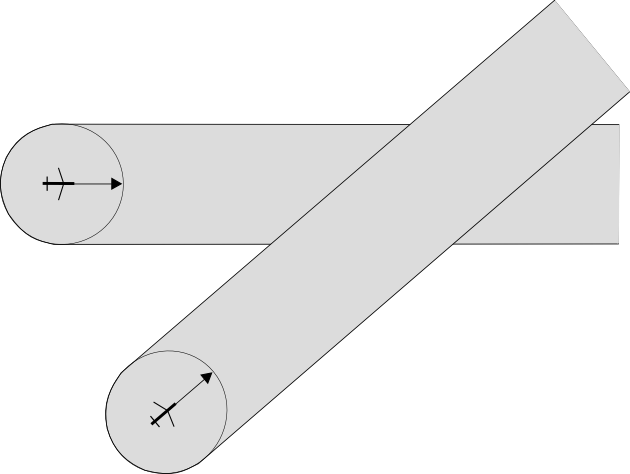}}
\subfigure[Growing uncertainties conflict avoidance]{\label{fig:growingConflict}
    \includegraphics[scale = 0.3]{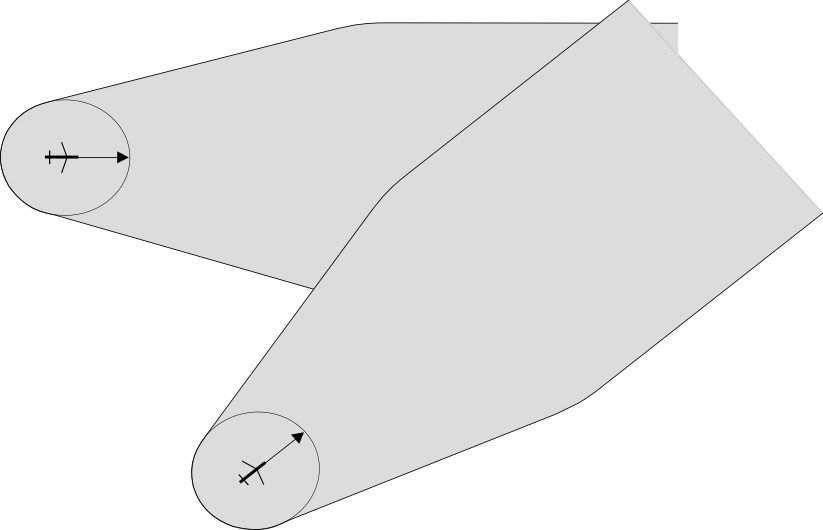}}
\caption{Conflict avoidance problems comparison}\label{fig:conflictComparison}
\end{center}
\end{figure}

From these considerations, we see it will not be difficult to create aircraft configurations which are conflict-free, yet will rapidly generate conflicts in case of positioning system degradation. Thus, the ability for such configurations to gracefully degrade will rely on our ability to develop a conflict resolution strategy for growing aircraft position uncertainties.  Figure \ref{fig:conflictComparison} shows the difference between a classical conflict avoidance problem and the problem we are solving. In the classical avoidance problem, the radius of avoidance is constant. In our problem, the growing radius of avoidance makes the formulation and resolution more complicated as it is time dependent.

When several aircraft are present within a given airspace sector, the conflict resolution problem under degrading position uncertainty becomes much more complex. An approach using mathematical programming based on a formulation originally presented in \cite{PallottinoConflitResolution} is introduced in Appendix. 
 The problem is formulated as a Mixed Integer Linear Program (MILP) and is implemented using the AMPL/CPLEX linear programming tool suite~\cite{AMPL,CPLEX}.

\section{Free Flight versus Miles-In-Trail analysis}
\begin{figure}[ht]
\centering
    \includegraphics[scale = 0.3]{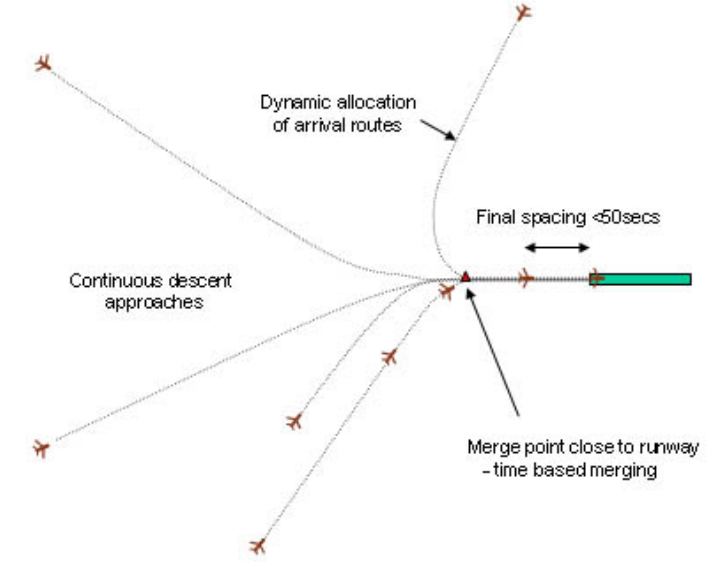}
    \caption{Arrival Routes, Multiple Merge Points \cite{SESARconceptOfOperation}}
\label{fig:configArrivalSESAR}
\end{figure}

The parts of airspace where the highest aircraft densities occur are the terminal areas in the vicinity of airports. Therefore, these constitute an ideal setting for evaluating the ability for traffic to undergo graceful degradation. 
This choice is also motivated by current vistas on future operations in terminal areas, which have been named ``Super-Density operations'' (NextGen) and ``High Complexity Terminal operations'' (SESAR). Both consist of increasing the airspace capacity around busy airports. A solution proposed in SESAR~\cite{SESARconceptOfOperation} is the multiple merge points arrival operation shown in Figure~\ref{fig:configArrivalSESAR}). One way to interpret this solution is to assimilate the new mode of operation to a ``Free-Flight'' scenario, whereby the route structure in the terminal area is relaxed and the only constraint on incoming aircraft is to meet a specific arrival time at the merge point. 
This scenario contrasts sharply with current airport arrival practices, where high-density arrival flows of aircraft are organized tens or even hundred of miles prior to landing, by lining up aircraft along arrival routes and spacing them appropriately. Such operations are often denoted Miles-In-Trail (MIT) operations.

Therefore we believe it is interesting and worthwhile to compare the ability for both operations, denoted Free-Flight (FF) and miles-in trail (MIT) to undergo graceful degradation of the navigation system.
For that purpose, we consider a simplified scenario that involve 8 aircraft merging to a common point of coordinates $(0,0)$. We assume that all aircraft are Time Based Separated (TBS): Each aircraft is given an arrival time so as to meet a precise and regular arrival rate at the merge point. Assuming all aircraft fly at the same speed of $200$kt, the aircraft must therefore be initially located on regularly spaced circles centered at the merge point. The inter-aircraft spacings are designed to emulate future Instrument Meteorological
Conditions (IMC) arrival operations on closely spaced parallel runways, such as San Francisco Airport: As observed in~\cite{GarielDynamicIOmodel}, the average interarrival time for each runway is slightly less than 2 minutes. This translates into a 3Nm average separation between aircraft when they fly at $200kt$ and the two runways are in use. This separation also turns out to be a {\em very} conservative estimate of separation requirements for satellite-based navigation systems \cite{powell2005}, leading us to an initial circle of avoidance of radius $r_0 = 1.5$NM as proposed by ICAO for ADS-B\cite{ICAOADSBseparationStandards}. The final radius of avoidance was chosen to be $r_f = 2.5$NM to reflect the surveillance degradation that would occur, should a GPS-based surveillance fail and backup radar-based technology had to be used. The rate of growth reflects a heading uncertainty $\Delta \theta = 5\dg$. Hence, $\dot r = 0.29$NM/min. The transition time between $r_0$ and $r_f$ is $T = 3.44$min.

The goal of this study is to evaluate the impact of changing arrival operations from Mile-In-Trail towards Free Flight. For that purpose, we consider an ``arrival cone'' whose vertex is at the merge point. When the cone's angular width is zero or takes small values, it corresponds to highly structured Miles-In-Trail operations. When the cone's angular width is large, it corresponds to less structured, Free-Flight-like operations. 

\begin{figure}[ht]
\centering
    \includegraphics[scale = 0.4]{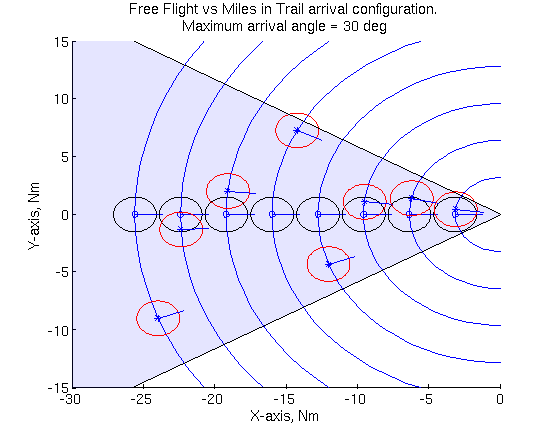}
    \caption{Free Flight and Miles-In-Trail arrival configurations}
    \label{fig:aircraftOnCircles}
\end{figure}

The remainder of our study consists of understanding the impact of the cone angular width and the aircraft initial separations on traffic degradation, should a surveillance system failure occur.
The cone angular width was varied from $10\dg$ to $60\dg$. 
The initial aircraft separation, that is, the distance between two consecutive circles, was chosen to be $3.5$NM. 1650 cases with different arrival angles have been simulated. The following procedure has been used to generate the cases: Since we know the distance of each aircraft to the merging point (fixed separation distance), the aircraft initial heading was picked at random, using a uniform probability distribution in the allowed interval ($\pm5\dg$, $\pm10\dg$, $\ldots$). Figure \ref{fig:aircraftOnCircles} presents two configurations: MIT and FF. The circles represent the avoidance circles (red = FF, black = MIT) and have initial radius $1.5$NM and are centered on the aircraft. The blue line shows aircraft's heading. Aircraft are represented by a small circle for the MIT configuration and asterisks for FF configuration. The allowed arrival cone for the FF configuration is depicted in light blue.

To analyze the impact of a surveillance system degradation, these traffic configurations have
been used as initial condition for the conflict resolution algorithm under uncertainties developed in appendix. Let $S$ be the index set of all generated cases. The severity of the traffic management degradation on the traffic situation $s\in S$
was evaluated by measuring the average deviation $m_s$ required for each aircraft, denoting $\theta_{is0}$ the initial heading of aircraft $i$ and $\theta_{i}$ its heading after resolution in this situation. 
\begin{equation}
 m_s = \frac{\sum_{i=1}^n |\theta_{is} - \theta_{is0}|}{n},
\end{equation}
where $n$ is the number of aircraft. All the cases were then sorted by absolute value of the maximum aircraft arrival angle and grouped in parameter increments of $2.5\dg$.

\begin{equation}
 \begin{split}
G_5 = \{ s \in S &\mbox{ such that } \max_i \{\theta_{is0}\} \in [0, 5), i=1\ldots n\},\\
G_k = \{ s \in S &\mbox{ such that } \max_i \{\theta_{is0}\} \in [k-1, k),\ldots\\
 &i=1\ldots n\}, k=7.5,10, 12.5,\ldots, 30.
 \end{split}
\end{equation}

\begin{figure}[ht]
\centering
    \includegraphics[scale = 0.4]{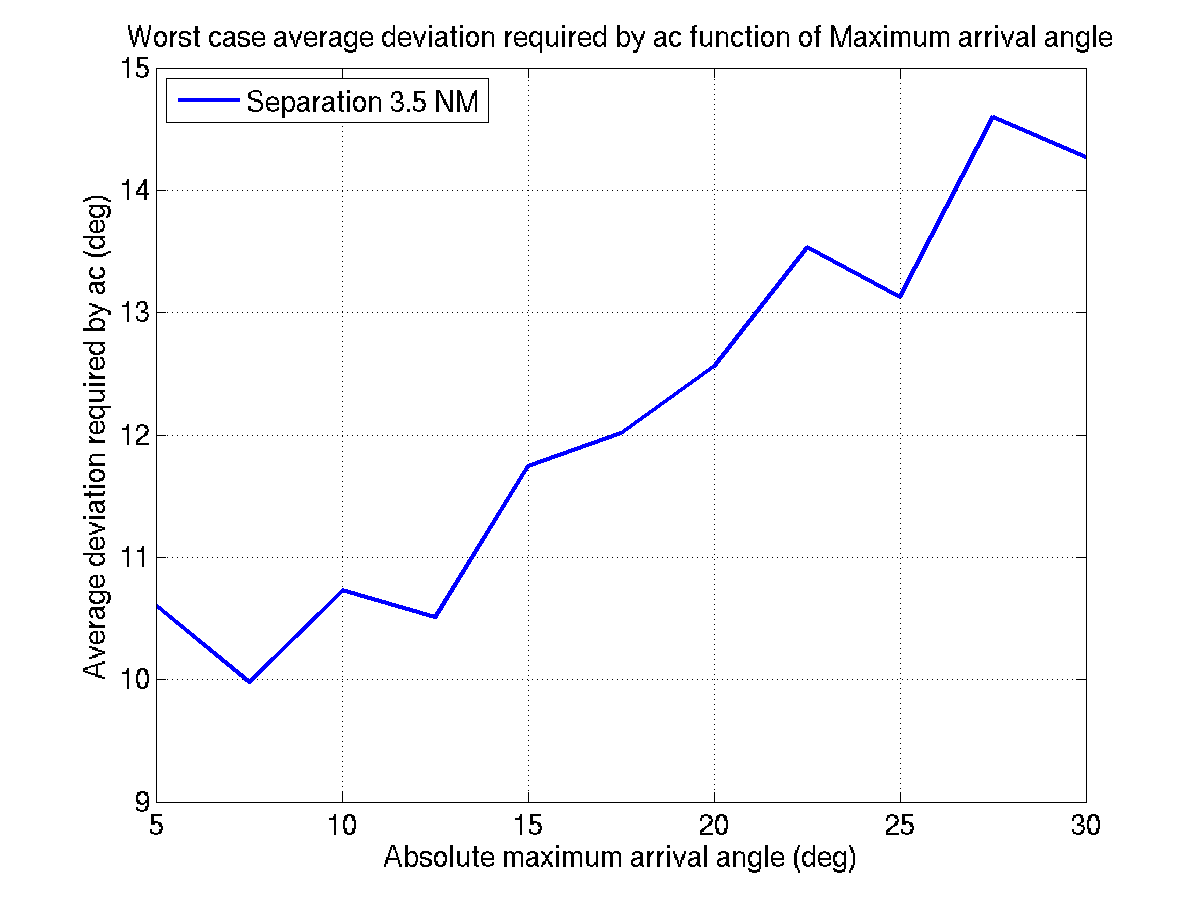}
    \caption{Evolution of the worst deviation required.}
    \label{fig:EvolutionDeviation3.5}
\end{figure}

Figure~\ref{fig:EvolutionDeviation3.5} presents the results of the analysis. The maximum deviation $m_s$ experienced within each group was plotted as a function of the corresponding maximum aircraft arrival angle.

\begin{figure}[ht]
\centering
\subfigure[$t=0min$]{\label{fig:worstG5_3.5init}
   \includegraphics[scale = 0.30]{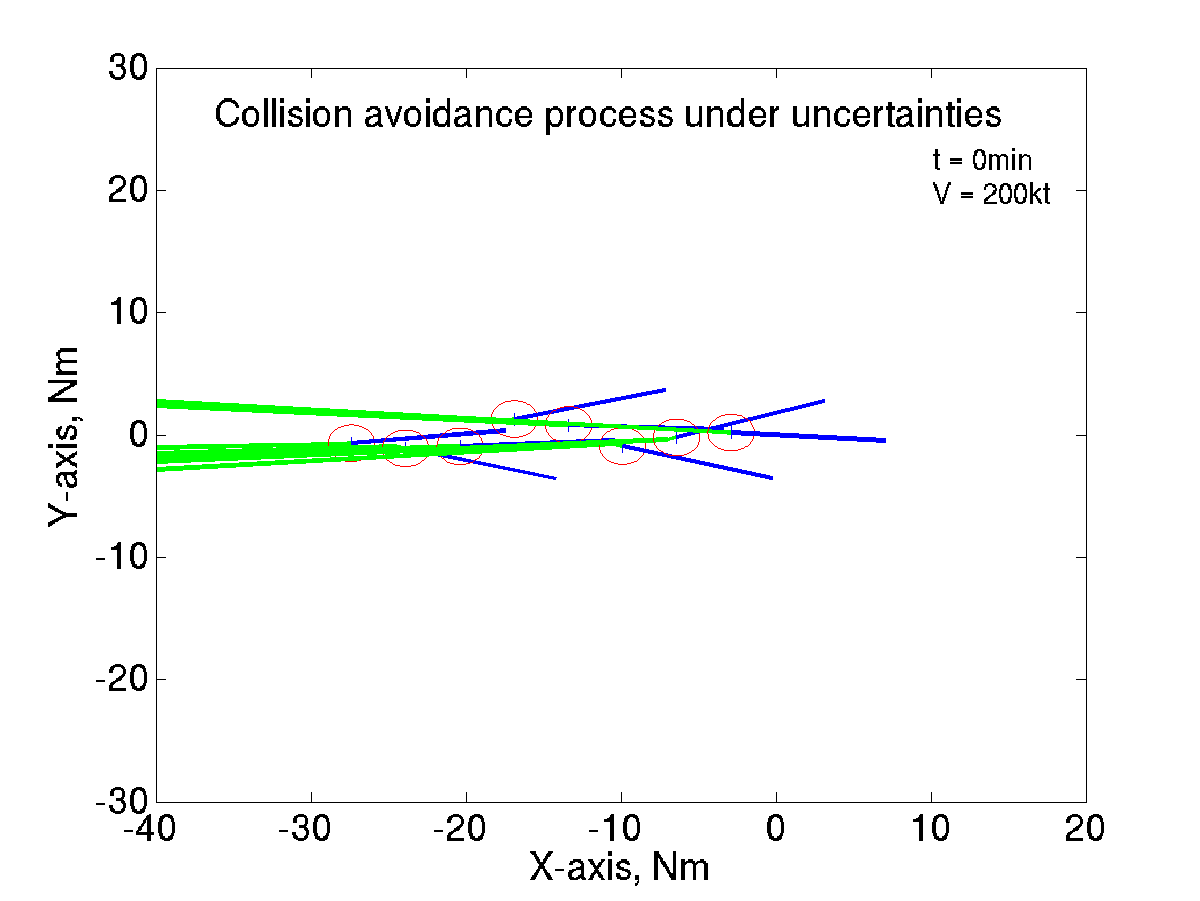}}
\subfigure[$t=3min$]{\label{fig:worstG5_3.5}
    \includegraphics[scale = 0.30]{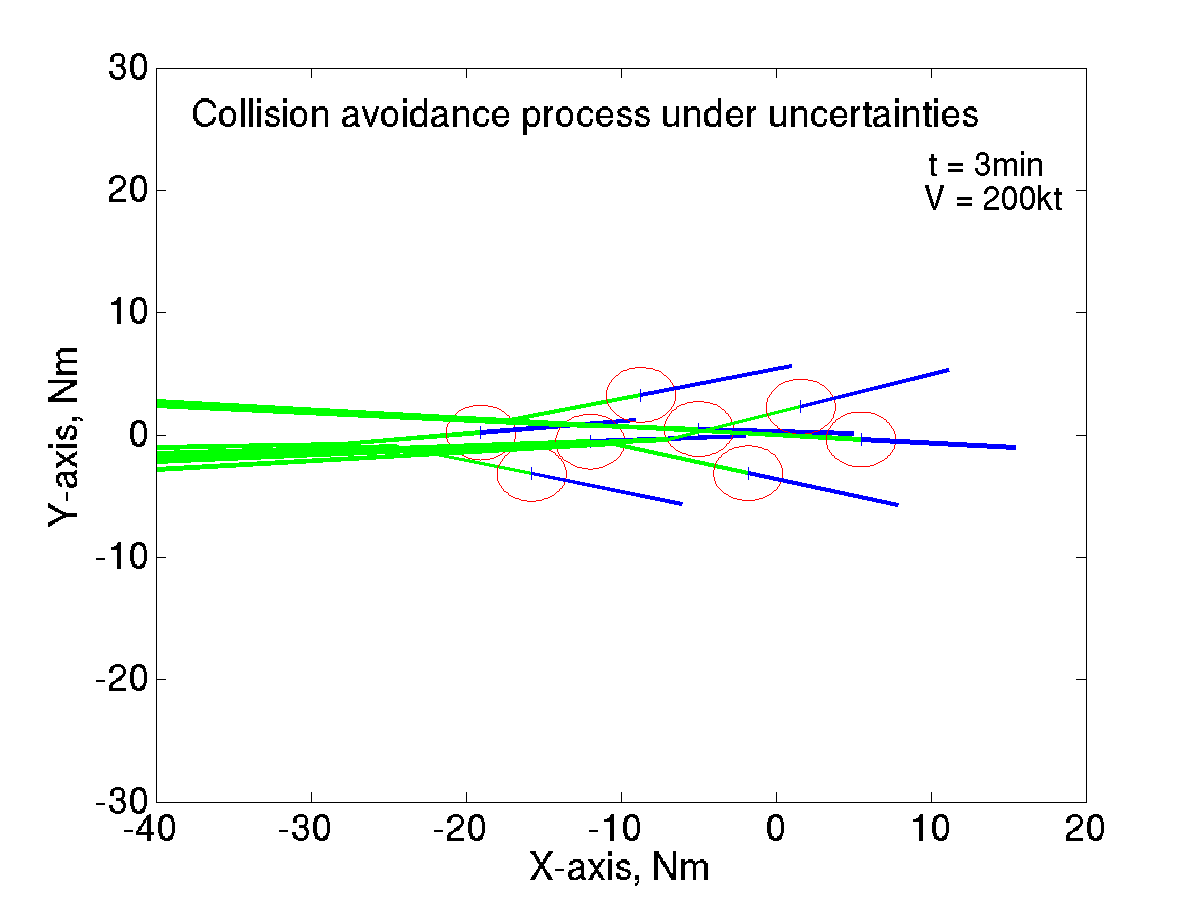}}
\caption{Avoidance maneuvers for the worst case of $G_5$}\label{fig:worstG5}
\end{figure}

\begin{figure}[ht]
\centering
\subfigure[$t=0min$]{\label{fig:worstG30_3.5init}
   \includegraphics[scale = 0.30]{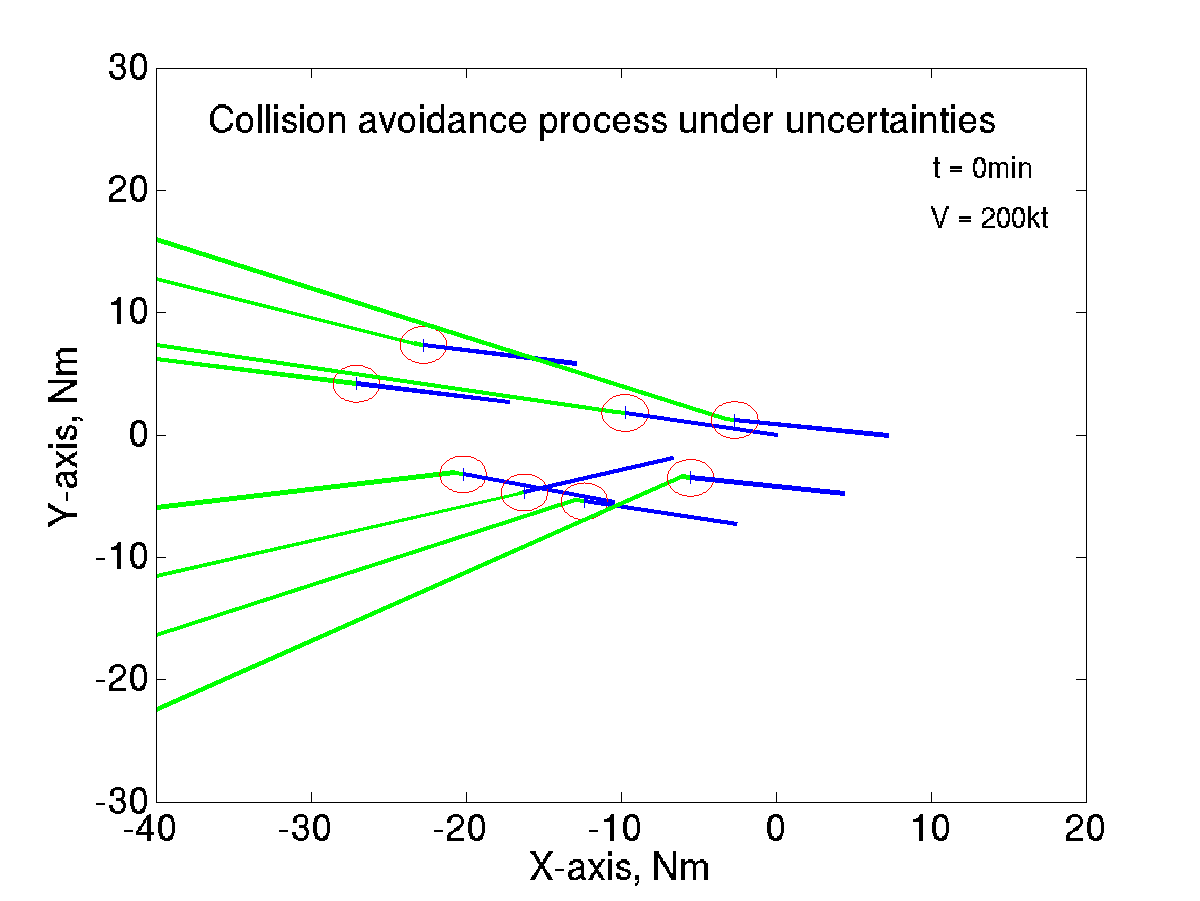}}
\subfigure[$t=3min$]{\label{fig:worstG30_3.5}
    \includegraphics[scale = 0.30]{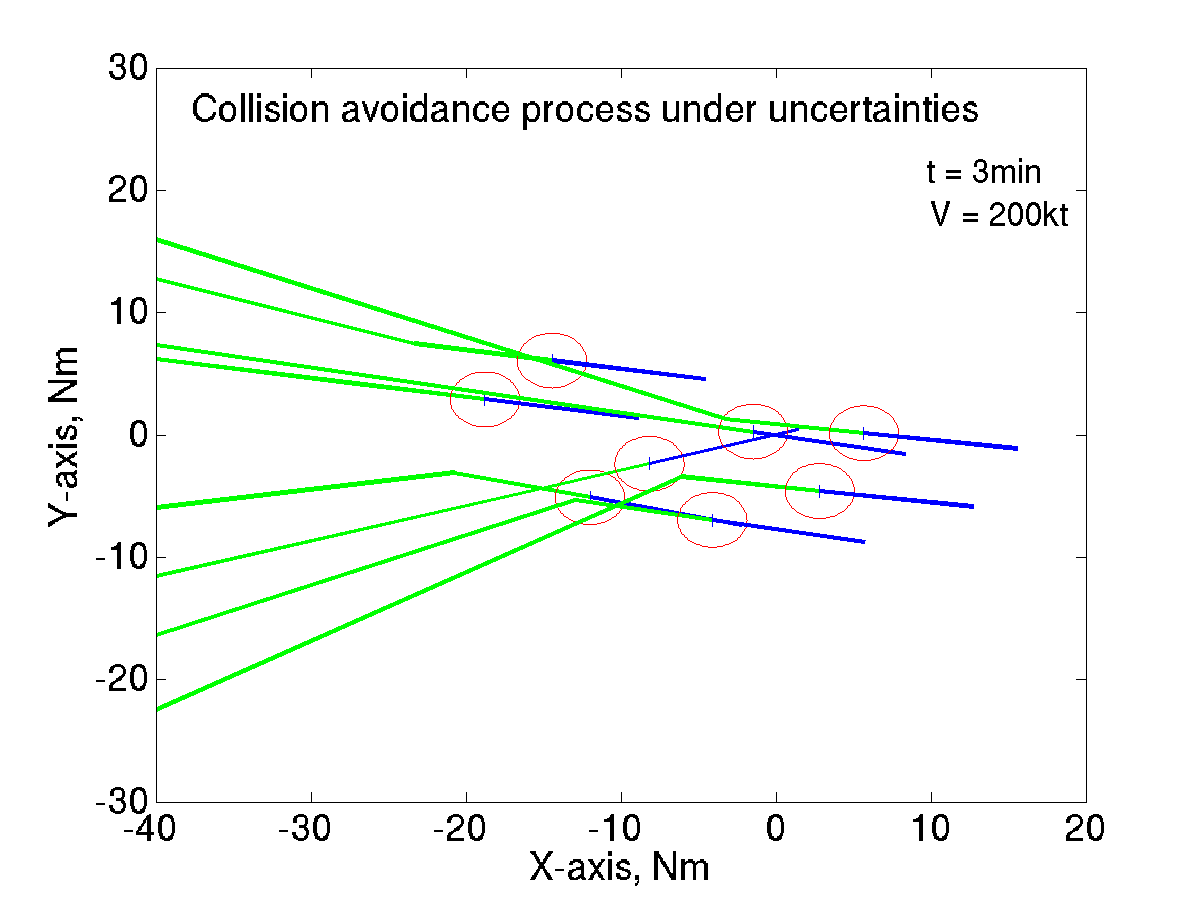}}
\caption{Avoidance maneuvers for the worst case of $G_{30}$}\label{fig:worstG30}
\end{figure}

Although the figure suffers from sampling irregularities, the following general trend may be observed: As the arrival cone increases, the aircraft deviations required to ensure a conflict free configuration increases 33\%. For an arrival angle less than $5\dg$, the maximum deviation required is  $10.6\dg$ per aircraft, while it is   $14.2\dg$ for a maximum arrival angle less than $30\dg$. 
Figures \ref{fig:worstG5} and \ref{fig:worstG30}  show the avoidance maneuvers for the worst cases of $G_5$ and $G_{30}$  with 3.5NM initial separation. Figure \ref{fig:worstG5_3.5init} and \ref{fig:worstG30_3.5init} present the configuration at $t=0$. Aircraft are at the postion where the avoidance maneuver is calculated. The trajectories of $t<0$ are represented and the line pointing out of the aircraft represent the new aircraft's heading.

 Free Flight or Miles-In-Trail do not appear to be significantly different from the stand point of surveillance degradation. However, these conclusions were reached using computer-based conflict management which may differ from human-based conflict management.

\section{Conclusion}
This paper has outlined the principle of graceful degradation of air traffic
operations in the face of Communication, Navigation or Surveillance system
failures.
Following a generic description of graceful degradation requirements and
interpreting it in the context of safety, we have introduced a specific
problem
of graceful degradation that examines the impact of failures of the
surveillance system on airborne traffic separation assurance. Considering
current Miles-In-Trail and
future Free-Flight approach scenarios, we have shown that free-flight-like
airport approaches
do not degrade significantly more than current Miles-In-Trail scenarios
when facing failures of the
surveillance system. During the process of this study, we have developed a
new conflict resolution tool that
applies to the transient conditions encountered during failures of the
surveillance system.

 \renewcommand{\theequation}{A-\arabic{equation}}
  \setcounter{equation}{0}  
\renewcommand{\thefigure}{A-\arabic{figure}}
  \setcounter{figure}{0}  
\renewcommand{\thetable}{A-\arabic{table}}
  \setcounter{table}{0}  

\appendix[An algorithm for conflict resolution under uncertainties]

The following appendix presents the algorithm used to solve the problem of conflict resolution when several aircraft are present. We decided to formulate the problem as a Mixed Integer Linear Program (MILP) as it is an efficient way to solve optimization problems. 
A typical MILP looks like:
\begin{align}
 \min_{x,z} \quad &f_1^T x + f_2^Tz\\
\text{subject to} \quad&A_1x + A_2z \leq b\\
\end{align}
where $f_1 \in \mathbb{R}^{m}, f_2 \in \mathbb{R}^{n} $, $x \in \mathbb{R}^{m}, z \in \{0,1\}^n $.  $A_1 \in \mathbb{R}^{m \times m}, A_2 \in \mathbb{R}^{n \times n} $, $b  \in \mathbb{R}^{m+n}$. In what follows we will focus our attention on developing a MILP model for the conflict resolution problem of interest in this paper.

We consider a set of $n$ aircraft in a planar space. Each aircraft $ac_i, i=1,\cdots,n$ is defined by its position $(x_i,y_i)$, its heading $\theta_i$ and its  speed $V_i$.

\begin{figure}[ht]
\centering
 \includegraphics[scale = 0.4]{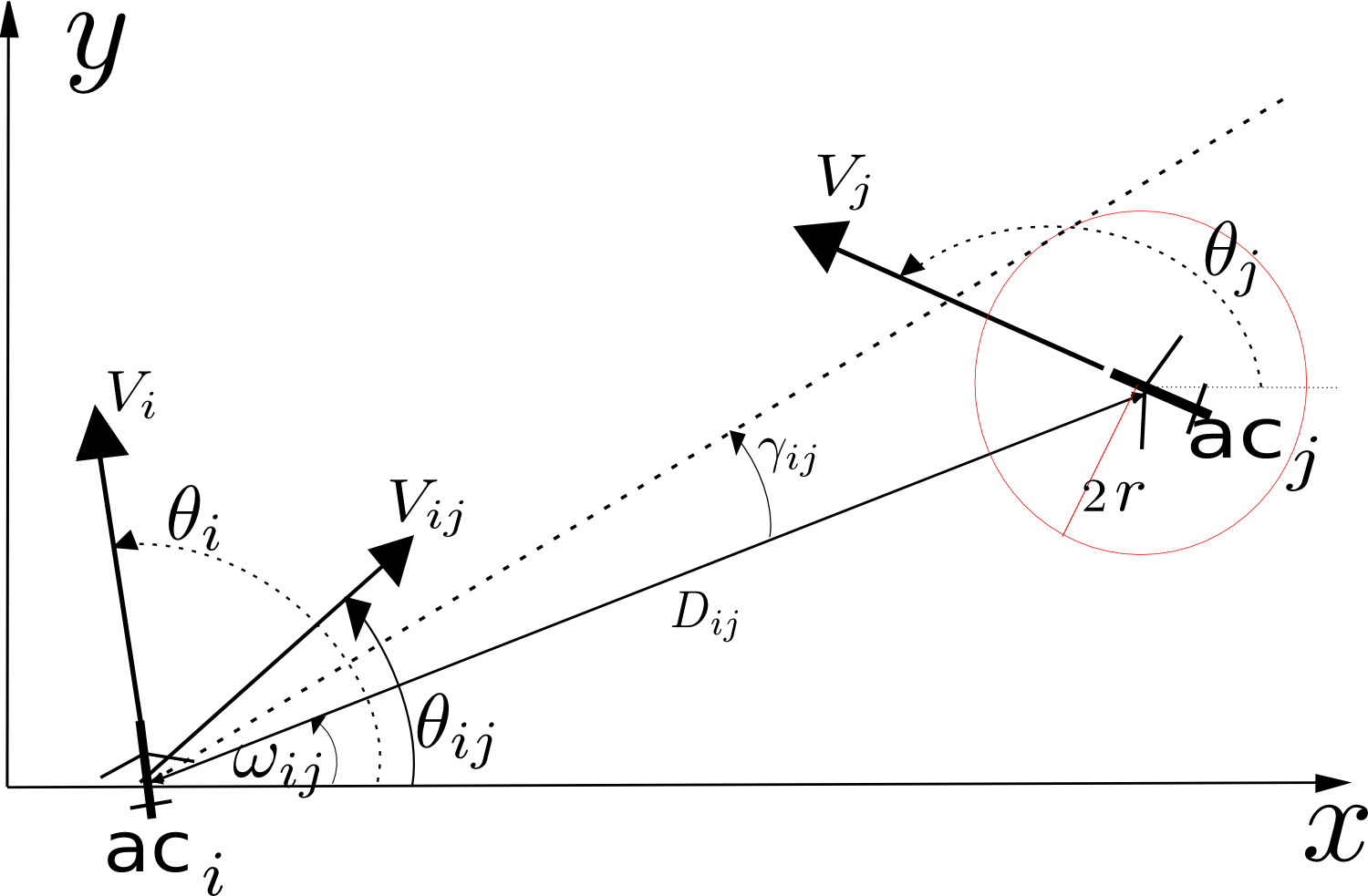}
\caption{Problem configuration}\label{fig:configuration}
\end{figure}

The relative velocity $\bV_{ij}$  and speed $V_{ij}$  of aircraft $i$ with respect to aircraft $j$, and the distance $D_{ij}$ between aircraft are given by:
\begin{align}
\bV_{ij} &= [(V_{x_i} - V_{x_j}), (V_{y_i} - V_{y_j})]^T\\
&= [V_i\cos\theta_i - V_j\cos\theta_j, V_i\sin\theta_i - V_j\sin\theta_j]^T,\\
V_{ij} &=\sqrt{(V_i\cos\theta_i{-}V_j\cos\theta_j)^2{+}(V_i\sin\theta_i{-}V_j\sin\theta_j)^2},\\
D_{ij} &= \sqrt{(x_i - x_j)^2 + (y_i - y_j)^2}.
\end{align}
See also Figure \ref{fig:configuration}. Let us define some useful parameters for the avoidance problem.
Let $\theta_{ij}$ be the angle between the relative velocity $\bV_{ij}$ and the $x$-axis, $\omega_{ij}$ be the angle between the connector of the aircraft and the $x$-axis. Finally, let $ \gamma_{ij}$ be the angle between the connector of the aircraft and a line starting from aircraft $ac_i$ and tangent to a circle of radius $2r$ and centered at aircraft $ac_j$. We have:
\begin{align}
\theta_{ij} &= \arctan{\frac{V_{y_{ij}}}{V_{x_{ij}}}}\\ \label{eq:thetaij}
&= \arctan{\frac{V_i\sin\theta_i - V_j\sin\theta_j}{V_i\cos\theta_i - V_j\cos\theta_j}}\\
\omega_{ij} &= \arctan{\frac{y_{j}-y_{i}}{x_{j}-x_{i}}}\\
\gamma_{ij} &= \arcsin{\frac{2r}{D_{ij}}},
\end{align}

\subsection{Problem structure}
\begin{figure}[ht]
\centering
 \includegraphics[scale = 0.30]{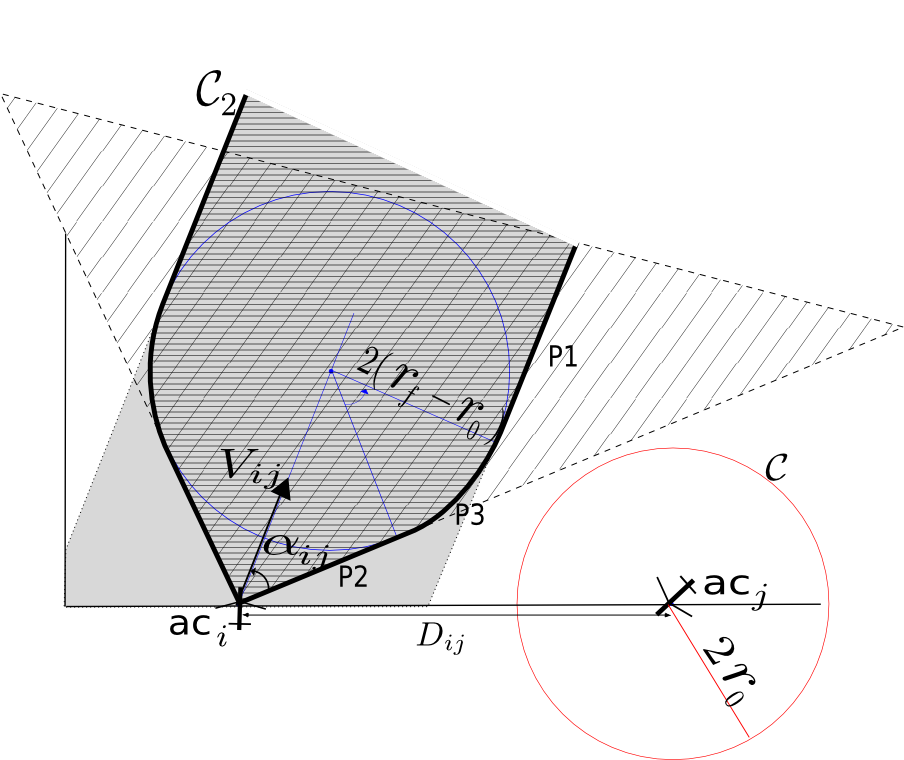}
\caption{Avoidance constraints}\label{fig:algoConstraints}
\end{figure}

We propose to solve the conflict resolution problem arising in this paper using a single heading change.  The originality of our problem lies with the fact that the allowable  miss distance between the two aircraft is time-dependent. Namely, at time $t=0$, the minimum miss distance is $2r_0$. For $0 \leq t \leq T$, the minimum miss distance grows from $2r_0$ to $2r_f$. $T$ is given by
\begin{equation}
T = \frac{r_f-r_0}{\dot r},
\end{equation}
where $\dot r$ is the growing rate of the circle of avoidance. For $t \geq T$, the miss distance is constant and equal to $2r_f$. Figure  \ref{fig:algoConstraints} illustrates the conflict avoidance constraint in a relative frame of reference. For no conflict to occur, the circle $\mathcal{C}$ of radius $2r_0$ and centered on aircraft $ac_j$ must not intersect the area enclosed by the contour $\mathcal{C}_2$. This contour $\mathcal{C}_2$ can be seen to be the union of the half line $P1$, the circular segment $P3$ and the line segment $P2$ and their symmetric images across the line passing though $ac_i$ and parallel to the velocity $\bV_{ij}$. 

For the purpose of linearization, we approximate the countour $\mathcal{C}_2$ by means of line segments, not to be intersected by the circle $\mathcal{C}$.  The first, obvious linear approximation is to ask that the circle $\mathcal{C}$ not intersect either the gray stripe in figure \ref{fig:algoConstraints}, or the hatched cone whose vertex is $ac_i$. 

\begin{itemize}
 \item Asking that the circle $\mathcal{C}$ not intersect the gray stripe is a classical conflict avoidance problem that was developed and solved by Pallottino in \cite{PallottinoConflitResolution}. This problem deals for time $t \geq T$, when enough time has elapsed for the two radii of avoidance to be equal to $r_f$. 
The avoidance problem presented in figure \ref{fig:algoConstraints} consists of finding a change in heading for aircraft $i$ and $j$ such that the line parallel to $\bV_{ij}$ and tangent to the circle of radius $2(r_f-r_0)$ centered on the relative position of aircraft $i$ to aircraft $j$ at $T$, does not intersect the circle of radius $2r_0$ centered on $ac_j$. 
This problem is equivalent to  the line directed along $\bV_{ij}$ and passing by aircraft $i$ does not intersect a circle of radius $2r_f$ and centered on $ac_j$. 
The avoidance constraints are then:
 \begin{equation}
\left\{
\begin{split}
&\theta_{ij} - \omega_{ij} > \tilde{\gamma}_{ij}\\
&\text{or}\\
&\theta_{ij} - \omega_{ij} < -\tilde{\gamma}_{ij}\\
\end{split}
\right.\\
\end{equation}
where 
\begin{equation}\label{eq:gammaTilde}
 \tilde{\gamma}_{ij} = \arcsin{\frac{2r_f}{D_{ij}}},
\end{equation}

\item Asking that the circle $\mathcal{C}$  not intersect the hatched cone can be detailed as follows, referring back to figure \ref{fig:algoConstraints}. The angular width $2\alpha_{ij}$ of the cone depends on the relative velocity of the aircraft: 
\begin{equation}
\sin\alpha_{ij} = \frac{\dot r_i + \dot r_j}{V_{ij}}.
\end{equation}
If ${\dot r_i + \dot r_j} > V_{ij}$, the sum of the radii of the avoidance circles increases faster than the aircraft go away from each other. Whatever the relative velocity, the circles are bound to intersect each other. Hence, the cone of avoidance is the entire plan: as $\arcsin\alpha_{ij}$ is not defined, we set $\alpha_{ij} = \pi$.  If ${\dot r_i + \dot r_j} = V_{ij}$, the increase rate of the circle of avoidance is the same as the relative speed. Hence, the avoidance cone is a half plane perpendicular to the relative velocity, $\alpha_{ij} = \frac{\pi}{2}$. The distance between the circles of avoidance will remain the same.
\begin{figure}[ht]
\centering
    \includegraphics[scale=0.3]{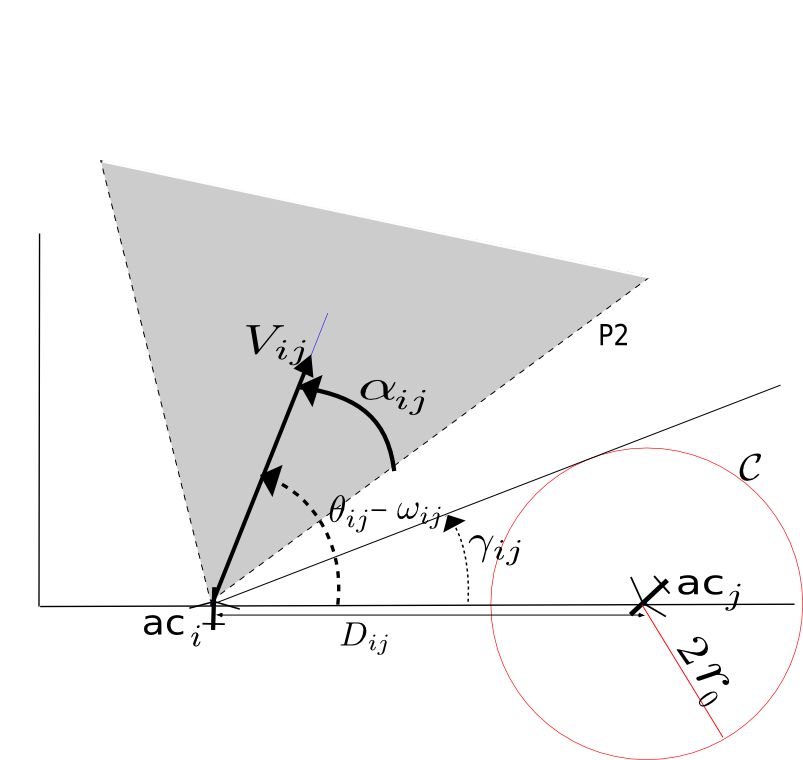}
    \caption{Configuration in the relative frame of reference: cone avoidance}
\label{fig:coneAvoidance}
\end{figure}
The condition of avoidance between two aircraft is given by a condition on angles, for $-\frac{\pi}{2} \leq \omega_{ij} \leq \frac{\pi}{2}$:

\begin{equation}
\left\{
\begin{split}
&\theta_{ij} - \omega_{ij} - \alpha_{ij} > \hat{\gamma}_{ij}\\
&\text{or}\\
&\theta_{ij} - \omega_{ij} + \alpha_{ij} < - \hat{\gamma}_{ij}\\
\end{split}
\right.\\
\end{equation}
where
\begin{equation}\label{eq:gammaHat}
 \hat{\gamma}_{ij} = \arcsin{\frac{2r_0}{D_{ij}}},
\end{equation}
Figure \ref{fig:coneAvoidance} presents those avoidance constraints. To avoid any singularity due to angles around $\pm \pi$, we ensure that $-\frac{\pi}{2} \leq \omega_{ij} \leq \frac{\pi}{2}$. To do so, aircraft are ordered in function of their position $(x_i,y_i)$ so that we get: $x_1 \leq x_2 \leq \ldots \leq x_n$ and if $x_i = x_{i+1}$, $y_i < y_{i+1}$.
\end{itemize}

\begin{figure}[ht]
\centering
    \includegraphics[scale=0.5]{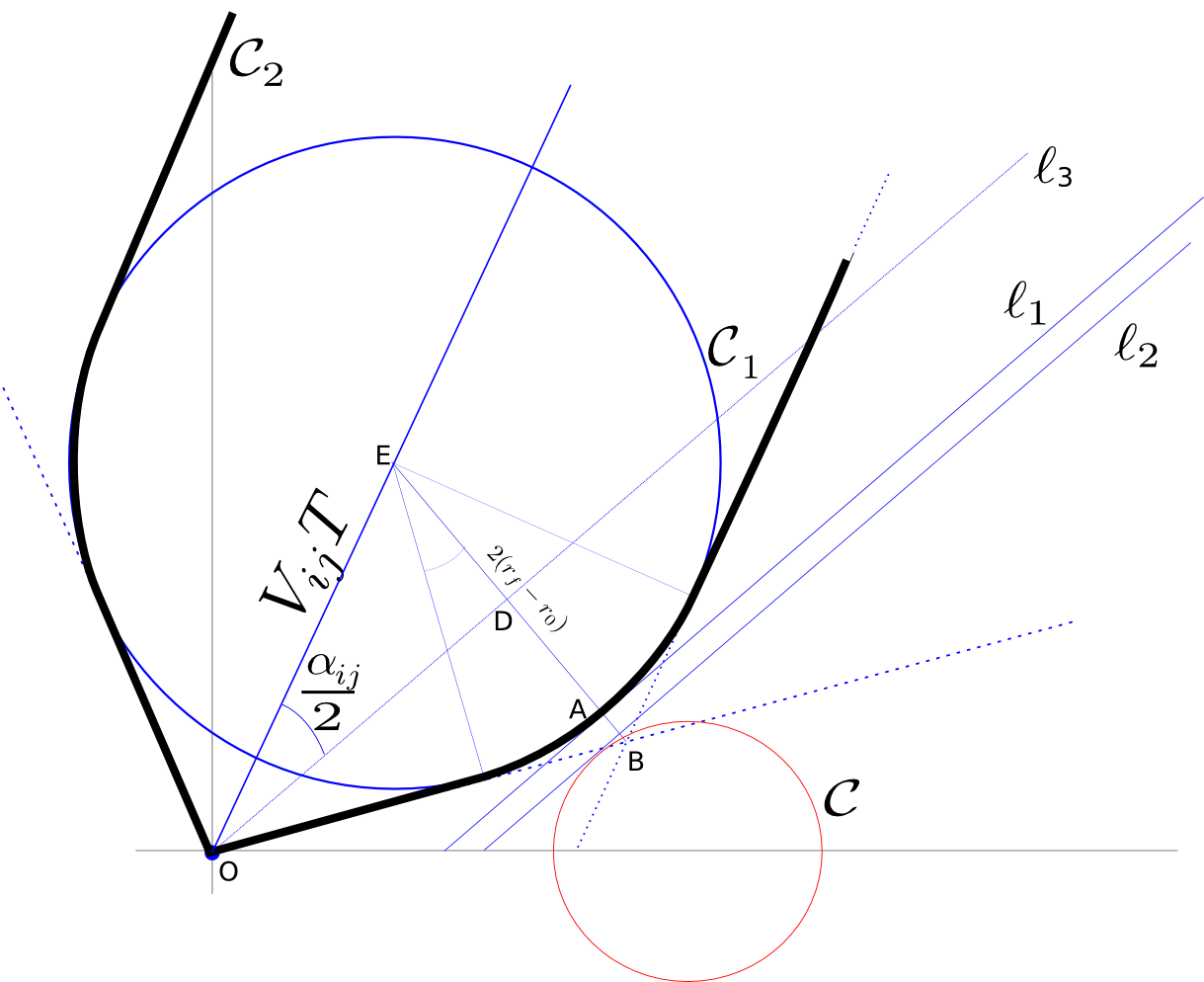}
    \caption{Geometry of the curved-part constraint (P3)}
\label{fig:3rdConstraint}
\end{figure}

The description of the avoidance constraints could be left at that point. However, the developed constraints are somewhat conservative. For example, as shown in figure \ref{fig:3rdConstraint}, the circle $\mathcal{C}$ may intersect the cone and the gray stripe without intersecting $\mathcal{C}_2$. To improve the solution, we may introduce a new constraint. Ideally, we would like to introduce the tangent ($\ell_1$) to $\mathcal{C}_2$ at point A. In the relative frame of reference, $\ell_1$ has a slope  $\theta_{ij} - \frac{\alpha_{ij}}{2} $.  $\ell_1$  not intersecting $\mathcal{C}_2$ is equivalent to $\ell_3$ not intersecting the circle centered on aircraft $j$ and of radius ($2r_0 + |AD|$). To formulate the constraint, we need the distance $|AD|$, between points $A$ and $D$. This distance is $2(r_f - r_0) - V_{ij}T\sin{\frac{\alpha{ij}}{2}}$. This distance is a non linear function of $\theta_i$ and $\theta_j$ as $V_{ij}$ and $\alpha_{ij}$ are non linear with respect to  $\theta_i$ and $\theta_j$. This function depends on too many parameters to be linearized easily. Nevertheless, we can approximate this constraint by using a line $\ell_2$: the line of slope $\theta_{ij} - \frac{\alpha_{ij}}{2} $ passing through a point between $A$ and $B$. For that, we need to find a majorant to $|DA|$.  The following geometrical development gives this majorant: $D$ is the middle of $EB$ and $A \in [DB]$.  
\begin{align}
ED&=\frac{EB}{2}\\
	&>\frac{EA}{2}\\
	&=r_f-r_0.
\end{align}
Hence, we get
\begin{align}
DA&=EA - ED\\
	&<r_f-r_0.
\end{align}

This leads to the constraint of the line of slope $\tfrac{\alpha{ij}}{2}$ not intersecting the circle of radius  $2r_0 + \tfrac{r_f-r_0}{2}= r_0 + r_f$: 

\begin{equation}
	\left\{
	\begin{split}
	&\theta_{ij} - \omega_{ij} - \frac{\alpha_{ij}}{2} > \gamma^{\ast}_{ij}\\
	&\text{or}\\
	&\theta_{ij} - \omega_{ij} + \frac{\alpha_{ij}}{2} < -\gamma^{\ast}_{ij}\\
	\end{split}
 	\right.\\
\end{equation}

where $\gamma^{\ast}_{ij}$ is given by
\begin{equation}\label{eq:gammaAst}
	\gamma^{\ast}_{ij} 	=  \arcsin\frac{r_0 + r_f}{D_{ij}}.
\end{equation}
 
As visible on figure \ref{fig:3rdConstraint}, this constraint improves the solution by allowing some solutions where the circle $\mathcal{C}$ was intersecting the first two constraints.

\subsection{Optimization of the conflict resolution}
In the previous section, we have presented constraints that should be satisfied by the aircraft headings to avoid a conflict. These constraints may be incorporated in an optimal conflict resolution scheme. Denoting the initial heading $\theta_{i0}$ of the aircraft $i$ and $\theta_i$ its heading after resolution, the problem is to compute:

\begin{equation}\label{eq:min}
\min J(\theta_1,\theta_2,\ldots \theta_n) = \min\sum_{i=1\ldots n}|\theta_i - \theta_{0i}|,
\end{equation}

subject to\newline
For all aircraft $i=1\ldots n-1$,\newline
\hspace{1cm}For all aircraft $j=i+1\ldots n$

\begin{equation}
\begin{split}
&\left\{
	\begin{split}
	&\theta_{ij} - \omega_{ij} > \tilde{\gamma}_{ij}\\
	&\text{and}\\
	&\theta_{ij} - \omega_{ij} < -\tilde{\gamma}_{ij}\\
	\end{split} 
\right.\\
\text{or}\\
&\left\{
	\begin{split}
	&\theta_{ij} - \omega_{ij} - \alpha_{ij} > \hat{\gamma}_{ij}\\
	&\text{and}\\
	&\theta_{ij} - \omega_{ij} + \alpha_{ij} < - \hat{\gamma}_{ij}\\
	\end{split}
\right.\\
\text{or}\\
&\left\{
	\begin{split}
	&\theta_{ij} - \omega_{ij} - \frac{\alpha_{ij}}{2} > \gamma^{\ast}_{ij}\\
	&\text{and}\\
	&\theta_{ij} - \omega_{ij} + \frac{\alpha_{ij}}{2} < -\gamma^{\ast}_{ij}\\
	\end{split}
 	\right.\\
\end{split}
\label{eq:generalAvoidanceConstraints}
\end{equation}

where $ \tilde{\gamma}_{ij}$, $\hat{\gamma}_{ij}$ and $ \gamma^{\ast}_{ij}$, are given by equations \ref{eq:gammaTilde}, \ref{eq:gammaHat} and \ref{eq:gammaAst}, respectively.

\subsection{Identical speeds case}
For the sake of computing simplicity, we assume all aircraft share the same speed $V$. This assumption simplifies the formulae for $\theta_{ij}$ and $\alpha_{ij}$ and allows us to obtain piecewise linear approximations. 

\subsubsection{Expression of the angle of the relative velocity: $\theta_{ij}$}
Using the assumption of identical speed, the angle between the relative velocity and the $x$-axis, the expression of $\theta_{ij}$ given by equation \ref{eq:thetaij} can be simplified : 
\begin{equation}
\begin{split}
\theta_{ij} &= \arctan\frac{\sin\theta_i - \sin\theta_j}{\cos\theta_i - \cos\theta_j}\\
&=\arctan\frac{\sin(\frac{\theta_i{-}\theta_j}{2})\cos(\frac{\theta_i{+}\theta_j}{2})}{-\sin(\frac{\theta_i{-} \theta_j}{2})\sin(\frac{\theta_i{+}\theta_j}{2})}.\\
\end{split}
\label{eq:thetaijLinear}
\end{equation}
$\theta_{ij}$ is a function of $\theta_i + \theta_j$ and $\theta_i - \theta_j$.  It can be shown that $\theta_{ij}$ is a piecewise affine function of $\theta_i + \theta_j$ and $\theta_i - \theta_j$ of the form $\theta_{ij} = m_{ij} (\theta_i + \theta_j) + p_{ij}$. The value of $m_{ij}$ if always $\tfrac{1}{2}$ and the values taken by $p_{ij}$  are summarized in table \ref{tab:ThetaijmCoef}.

\renewcommand{\arraystretch}{1.5}
\begin{table}[ht]
\begin{center}
    \begin{tabular}{|p{0.1\linewidth}|p{0.1\linewidth}|p{0.20\linewidth}|}
    \hline
    \multicolumn{1}{|c|}{Case 1}
    &\multicolumn{1}{|c|}{$\theta_i - \theta_j < 0$ and $-2\pi \leq \theta_i + \theta_j < -\pi$}
    &\multicolumn{1}{|l|}{$p_{ij} = \frac{3\pi}{2}$}\\
    \hline
    \multicolumn{1}{|c|}{Case 2}
    &\multicolumn{1}{|c|}{$\theta_i - \theta_j < 0$ and $-\pi \leq \theta_i + \theta_j < 2\pi$}
    &\multicolumn{1}{|l|}{$p_{ij} = -\frac{\pi}{2}$}\\
    \hline
    \multicolumn{1}{|c|}{Case 3}
    &\multicolumn{1}{|c|}{$\theta_i - \theta_j \geq 0$ and $-2\pi \leq \theta_i + \theta_j < \pi$}
    &\multicolumn{1}{|l|}{$p_{ij} = \frac{\pi}{2}$}\\
    \hline
    \multicolumn{1}{|c|}{Case 4}
    &\multicolumn{1}{|c|}{$\theta_i - \theta_j \geq 0$ and $\pi \leq \theta_i + \theta_j < 2\pi$}
    &\multicolumn{1}{|l|}{$p_{ij} = -\frac{3\pi}{2}$}\\
    \hline
    \end{tabular}\caption{Intercept coefficient of $\theta_{ij}$}\label{tab:ThetaijmCoef}
\end{center}
\end{table}

We can determine the value of $p_{ij}$ using boolean variables. Let $bCaseDiffPos_{ij}$, $ bCaseSumInfmPi_{ij}$, $bCaseSumSupPi_{ij}$, $bCase1_{ij}$ and $bCase4_{ij}$ be boolean variables such that:
\begin{align}
&bCaseDiffPos_{ij} =1 \Longleftrightarrow \theta_i - \theta_j \geq 0\\
&bCaseSumInfmPi_{ij} = 1 \Longleftrightarrow  -2\pi \leq \theta_i + \theta_j < -\pi\\
&bCaseSumSupPi_{ij} = 1 \Longleftrightarrow \pi \leq \theta_i + \theta_j < 2\pi\\
&bCase1_{ij}=1 \Longleftrightarrow
\begin{cases}
&bCaseDiffPos_{ij} = 0\\
&\text{and} \\
&bCaseSumInfmPi_{ij} = 1
\end{cases}\\
&bCase4_{ij}=1 \Longleftrightarrow
\begin{cases}
&bCaseDiffPos_{ij} = 1\\
&\text{and} \\
&bCaseSumSupPi_{ij} = 1\\
\end{cases}\\
\end{align}

Those 3 boolean variables yield the following expression for $p_{ij}$ and hence $\theta_{ij}$:
\begin{equation}
p_{ij}{=}(bCaseSumPos_{ij}-\tfrac{1}{2})\pi + 2\pi(bCase1_{ij} - bCase4_{ij}),\\
\end{equation}
\begin{multline}
 \theta_{ij} = \frac{\theta_i + \theta_j}{2} + (bCaseDiffPos_{ij}-\tfrac{1}{2})\pi + \ldots\\ 
\hspace{2cm} 2\pi(bCase1_{ij} - bCase4_{ij}),
\end{multline}
which is linear in $\theta_i$ and $\theta_j$ and in the boolean variables.

\subsubsection{Constraints formulation for $\theta_{ij}$}
The following formulation uses the big-$M$ method. $M$ is a large number such that if multiplied by a boolean set to 1, the constraint is always satisfied, whatever the value of the other variables. This enables us to select between constraints and use $or$ relationship between constraints. This is a standard procedure described  \cite{OperationsResearch}.
 
Determination of the sign of  $\theta_i - \theta_j$:
\begin{multline}
bCaseDiffPos_{ij} = 1 \Longleftrightarrow\\
\begin{cases}
\theta_i - \theta_j  - M bCaseDiffPos_{ij} < 0,\\
-\theta_i + \theta_j - M (1-bCaseDiffPos_{ij}) < 0.
\end{cases}\\
\end{multline}
Determination of the boolean $bCaseSumInfmPi_{ij}$ if $\theta_i + \theta_j < -\pi$:
\begin{multline}
bCaseSumInfmPi_{ij} = 1 \Longleftrightarrow\\
\begin{cases}
\theta_i + \theta_j + \pi - M (1-bCaseSumInfmPi_{ij}) < 0,\\
-\theta_i - \theta_j - \pi - M bCaseSumInfmPi_{ij} < 0.
\end{cases}\\
\end{multline}
Determination if $\theta_i + \theta_j \geq \pi$:
\begin{multline}
bCaseSumInfmPi_{ij} = 1 \Longleftrightarrow\\
\begin{cases}
-\theta_i - \theta_j + \pi - M (1-bCaseSumSupPi_{ij}) \leq 0,\\
\theta_i + \theta_j - \pi - M bCaseSumSupPi_{ij} \leq 0.
\end{cases}\\
\end{multline}
Determination of the boolean $bCase1_{ij}$:
\begin{multline}
bCase1_{ij} = 1 \Longleftrightarrow\\
\begin{cases}
-1.5 + bCaseDiffPos_{ij} - \ldots \\
\quad bCaseSumInfmPi_{ij} + 2bCase1_{ij} \leq 0,\\
-0.5 - 2bCaseDiffPos_{ij}  \ldots\\
 \quad  + bCaseSumInfmPi_{ij} - bCase1_{ij} \leq 0,.
\end{cases}\\
\end{multline}
Determination of the boolean $bCase4_{ij}$:
\begin{multline}
bCase4_{ij} = 1 \Longleftrightarrow\\
\begin{cases}
1.5 - bCaseDiffPos_{ij} - bCaseSumSupPi_{ij} \ldots \\
\quad- 2(1-bCase4_{ij}) \leq 0,\\
 -1.5 + bCaseDiffPos_{ij} + bCaseSumSupPi_{ij} \ldots \\
\quad- bCase4_{ij} \leq 0.
\end{cases}\\
\end{multline}

\subsubsection{Cone angular width $2\alpha_{ij}$ in the relative frame of reference}
This section presents the formulas to determine the cone angular width $2\alpha_{ij}$ when aircraft share an identical speed. The assumption leads to $\dot r_i = \dot r, i=1\ldots n$ and
\begin{equation}
\begin{split}
\alpha_{ij} &= \arcsin\frac{2\dot r}{\sqrt{V^2(\cos\theta_i{-}\cos\theta_j)^2{+}V^2(\sin\theta_i{-}\sin\theta_j)^2}}\\
&= \arcsin\frac{\dot r}{V|\sin(\frac{\theta_i-\theta_j}{2})|}\\
&= \arcsin\frac{\tan\Delta\theta}{|\sin(\frac{\theta_i-\theta_j}{2})|}
\end{split}
\label{eq:alpha}
\end{equation}

The half cone angular width $\alpha_{ij}$ is given by equation (\ref{eq:alpha}) and it is a non linear function of $\theta_i - \theta_j$.  This function is symmetric about $\theta_i - \theta_j = 0$ and consists of two quasi convex components, separated at $\theta_i - \theta_j = 0$, as shown in Figure \ref{fig:alphaLinear}. The epigraph of each of these quasi convex components can be approximated by the intersection of linear constraints defined by their
slopes $a_k$ and intercepts $b_k$. Using the big-$M$ method allows us to account for the presence of two disconnected components, and this linearization leads to:

\begin{figure}[ht]
\centering
    \includegraphics*[scale=0.40,viewport=75 0 700 315 ]{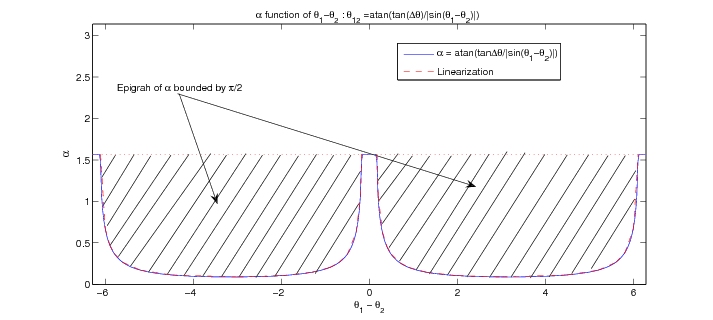}
    \caption{$\alpha_{ij}$ function of $\theta_i-\theta_j$ and the linearization used}
\label{fig:alphaLinear}
\end{figure}

\begin{equation}
\begin{split}
&a_1(\theta_i-\theta_j) + b_1 - M bCaseAlphaPos_{ij} \leq \alpha_{ij}\\
&a_2(\theta_i-\theta_j) + b_2 - M bCaseAlphaPos_{ij} \leq \alpha_{ij}\\
\vdots\\
&a_l(\theta_i-\theta_j) + b_l - M bCaseAlphaPos_{ij} \leq \alpha_{ij}\\
&-a_1(\theta_i-\theta_j) + b_1 - M(1 - bCaseAlphaPos_{ij}) \leq \alpha_{ij}\\
&-a_2(\theta_i-\theta_j) + b_2 - M(1 - bCaseAlphaPos_{ij}) \leq \alpha_{ij}\\
\vdots\\
&-a_l(\theta_i-\theta_j) + b_l - M(1 - bCaseAlphaPos_{ij}) \leq \alpha_{ij}
\end{split}
\end{equation}
with $bCaseAlphaPos_{ij}$ a binary variable used to transform the $or$ between the constraints in an $and$ relationship.

The avoidance constraints are now given by:
\begin{align}
\begin{cases}\label{eq:avoidanceConstraints}
    -\frac{\theta_i + \theta_j}{2} - (\frac{1}{2} - bCasePlus_{ij})\pi + \omega_{ij} + \alpha_{ij} \ldots\\
\quad + \hat\gamma_{ij} - M bCaseIneqPos_{ij} < 0\\
\text{and} \\
    \frac{\theta_i + \theta_j}{2} + (\frac{1}{2} - bCasePlus_{ij})\pi - \omega_{ij} + \alpha_{ij}  \ldots\\
\quad + \hat\gamma_{ij} - M(1 - bCaseIneqPos_{ij}) <0
\end{cases}
\end{align}
with $bCaseIneqPos_{ij}$ a binary variable used to transform the $or$ between the constraints in an $and$ relationship.

\subsubsection{Mixed constraints for conflict avoidance}
The conflict avoidance constraints given by equation \ref{eq:generalAvoidanceConstraints} can be handled in a linear programming framework by using the big-$M$ method, leading to the following mixed constraints: 

\begin{align}
\begin{cases}
    -\frac{\theta_i + \theta_j}{2} - (\frac{1}{2} - bCasePlus_{ij})\pi + \omega_{ij} + \tilde\gamma_{ijr_0} \ldots\\
\quad - M bCaseIneqPos_{ij} - M b1_{ij} < 0\\
    \frac{\theta_i + \theta_j}{2} + (\frac{1}{2} - bCasePlus_{ij})\pi - \omega_{ij} + \tilde\gamma_{ijr_0} \ldots\\ 
\quad - M(1 - bCaseIneqPos_{ij}) - M b1_{ij} <0\\
    -\frac{\theta_i + \theta_j}{2} - (\frac{1}{2} - bCasePlus_{ij})\pi + \omega_{ij} + \alpha_{ij} + \hat\gamma_{ijr_f} \ldots\\ \quad-  M bCaseIneqPos_{ij} - M b2_{ij} < 0\\
    \frac{\theta_i + \theta_j}{2} + (\frac{1}{2} - bCasePlus_{ij})\pi - \omega_{ij} + \alpha_{ij} + \hat\gamma_{ijr_f} \ldots\\ \quad- M(1 - bCaseIneqPos_{ij}) - M b2_{ij} <0\\
   -\frac{\theta_i + \theta_j}{2} - (\frac{1}{2} - bCasePlus_{ij})\pi + \omega_{ij} + \frac{\alpha_{ij}}{2} + \gamma^\ast_{ijr_f} \ldots\\ 
\quad- M bCaseIneqPos_{ij} - M b3_{ij}< 0\\
    \frac{\theta_i + \theta_j}{2} + (\frac{1}{2} - bCasePlus_{ij})\pi - \omega_{ij} + \frac{\alpha_{ij}}{2} + \gamma^\ast_{ijr_f}\ldots\\ \quad - M(1 - bCaseIneqPos_{ij}) - M b3_{ij} <0\\
b1_{ij} + b2_{ij} + b3_{ij} = 2\\
\end{cases}
\end{align}
with $b1_{ij}, b2_{ij}$ and $b3_{ij}$ binary variables. 

The global algorithm consists on minimizing expression \ref{eq:min} subject to all the constraints previously developed. We wrote the constraints in an AMPL format and solved using CPLEX. MATLAB was used to generate the 1650 cases and was also used for the post-processing. The computing time to solve an 8 aircraft configuration ranged from less than a second to dozen of seconds on a 4-processor, Pentium  class computer.

\section*{Acknowledgment}
This research was sponsored by Thales Air Systems. The authors would like to thank John Hansman from MIT for useful discussions about parallel approaches.

\bibliographystyle{unsrt}
\addcontentsline{toc}{section}{Bibliography}
\bibliography{ieeearticleMain}
\end{document}